\documentclass[11pt, oneside]{article}   	
\usepackage{geometry}                		
\usepackage{graphicx}				
\usepackage{amssymb}
\usepackage{graphicx}

\newcommand{\sext}{S_{\rm ext}}
\newcommand{\srich}{S_{\rm ext}^{\rm rich}}
\newcommand{\spoor}{S_{\rm ext}^{\rm poor}}
\newcommand{\tst}{T_{\rm stv}}
\newcommand{\tdec}{T_{\rm dec}}
\newcommand{\bvec}[1]{\mbox{\boldmath $#1$}}

\title{Theory for transitions between log and stationary phases: universal laws for lag time}

\author{Yusuke Himeoka\footnote{himeoka@complex.c.u-tokyo.ac.jp}, and Kunihiko Kaneko\footnote{kaneko@complex.c.u-tokyo.ac.jp}\\
Department of Basic Science, University of Tokyo\\ Komaba, Meguro-ku, Tokyo 153-8902, Japan}

\begin{document}
\maketitle

\begin{abstract}
Quantitative characterization of bacterial growth has gathered substantial attention since Monod's pioneering study. Theoretical and experimental work has uncovered several laws for describing the log growth phase, in which the number of cells grows exponentially. However, microorganism growth also exhibits lag, stationary, and death phases under starvation conditions, in which cell growth is highly suppressed, while quantitative laws or theories for such phases are underdeveloped. In fact, models commonly adopted for the log phase that consist of autocatalytic chemical components, including ribosomes, can only show exponential growth or decay in a population, and phases that halt growth are not realized. Here, we propose a simple, coarse-grained cell model that includes inhibitor molecule species in addition to the autocatalytic active protein. The inhibitor forms a complex with active proteins to suppress the catalytic process. Depending on the nutrient condition, the model exhibits the typical transition among the lag, log, stationary, and death phases. Furthermore, the lag time needed for growth recovery after starvation follows the square root of the starvation time and is inverse to the maximal growth rate, in agreement with experimental observations. Moreover, the distribution of lag time among cells shows an exponential tail, also consistent with experiments. Our theory further predicts strong dependence of lag time upon the speed of substrate depletion, which should be examined experimentally. The present model and theoretical analysis provide universal growth laws beyond the log phase, offering insight into how cells halt growth without entering the death phase.
\end{abstract}
~~~~Quantitative characterization of a cellular state, in terms of cellular growth rate, concentration of external resources, as well as abundances of specific components, has long been one of the major topics in cell biology, ever since the pioneering study by Monod \cite{monod1949growth}. Quantitative growth laws have been uncovered mainly by focusing on the microbial log phase in which the number of cells grows exponentially, including Pirt's equation for yield and growth \cite{Pirt224} and the relationship between the fraction of ribosomal abundance and growth rate (experimentally demonstrated by Schaechter {\it et al.}\cite{schaechter1958dependency}, and theoretically rationalized by Scott {\it et al.} \cite{scott2010interdependence}), among others \cite{bennett1974effects,ishii2007multiple,klumpp2009growth,madar2013promoter}, in which the constraint to maintain steady growth leads to general relationships\cite{furusawa2003zipf,himeoka2014entropy,kaneko2015universal}. In spite of the importance of the discovery of these universal laws, cells under poor conditions exhibit different growth phases in which such relationships are violated. Indeed, in addition to the death phase, cells undergo a stationary phase under conditions of resource limitation, in which growth is drastically suppressed. Once cells enter the stationary phase, a certain time span is generally required to recover growth after resources are supplied, which is known as the lag phase. Although several quantities have been measured to characterize these phases, such as the length of lag time for resurrection, and the tolerance time for starvation or antibiotics \cite{augustin2000model,gefen2014direct,vasi1999ecological}, there has been no theory put forward to characterize the phase changes, and no corresponding quantitative laws have been established.\\
~~~~To develop a theory for bacterial physiology beyond the log phase, we first constructed a simple mathematical model that exhibits the changes among the lag, log, stationary, and death phases. We then uncovered the quantitative characteristics of each of these phases in line with experimental observations. Including bacterial growth curve, quantitative relationships of lag-time with starvation time and the maximal growth rate, exponentially-tailed distribution of lag-time, and trade-off between the growth rate and tolerance for the starvation.  These are formulated by the changes in inhibitor (or mistranslated proteins) chemicals in addition to changes in ribosomal proteins (ribosomes). The proposed model also allowed us to reach several experimentally testable predictions, including the dependence of lag time on the speed of the starvation process.

\section*{Model}
\begin{figure}[h!]
\begin{center}
\includegraphics[width = 120 mm, angle = 0,bb=0 0 1345 1728]{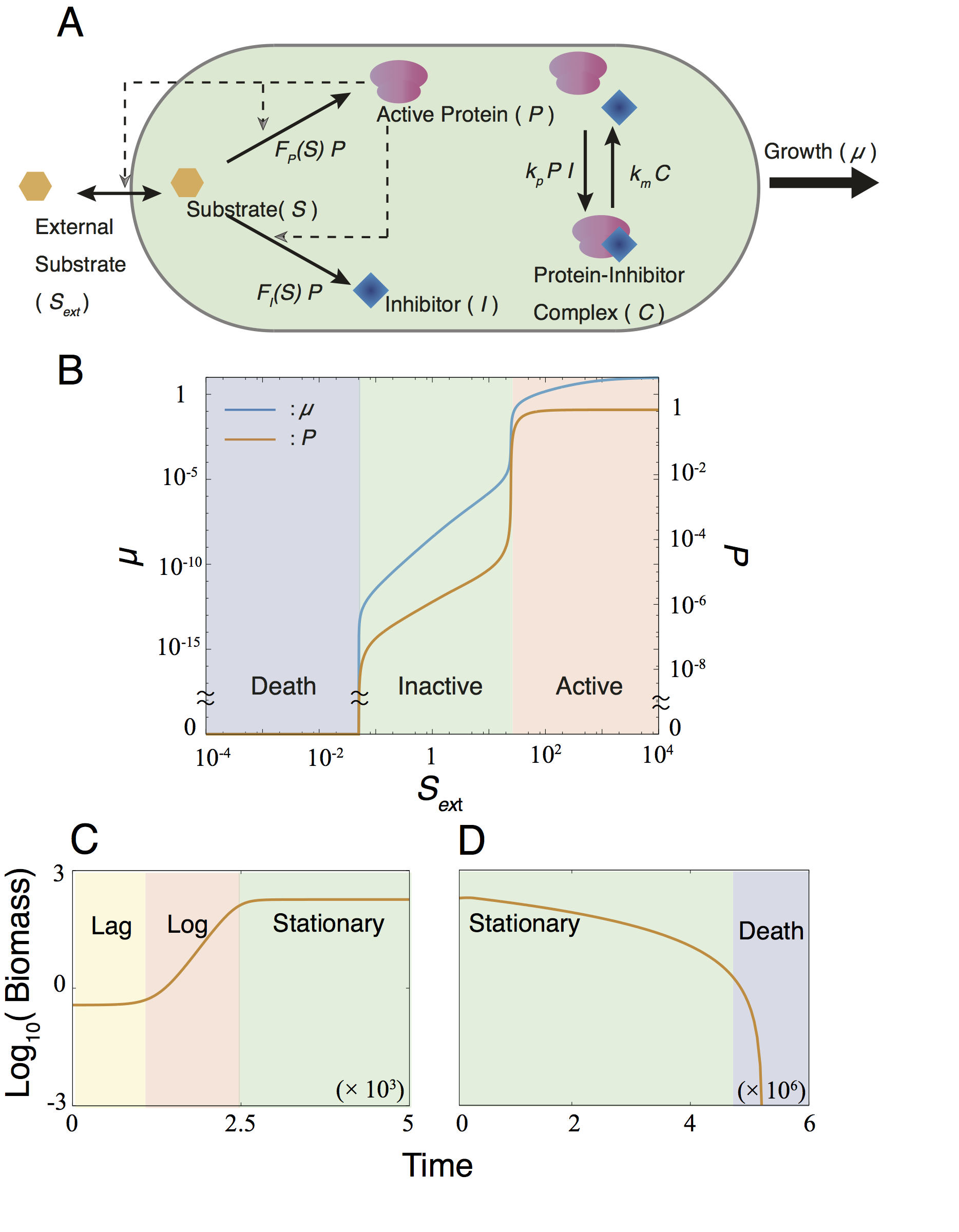}
\end{center}
     \caption{(A) Schematic representation of components and reactions in the present model. The concentration of each chemical changes according to the listed reactions. In addition, chemicals are spontaneously degraded at a low rate, and become diluted due to volume expansion of the cell. (B) Steady growth rate and the concentration of active protein are plotted as functions of the external concentration of the substrate.  (C and D) Growth curve of the model.  Parameters were set as follows: $v=10.0,k_p=0.02,k_m=2\times 10^{-8},K=5.0,K_t=100.0,d_R=d_I=10^{-5},d_C=10^{-7}$ for (B). Model and parameters for (C) and (D) are given in {\it Supplementary\ Information}. }
     
    \label{fig:Fig1}
\end{figure}

Models for growing cells generally consist of substrates($S$) and active proteins that catalyze their own synthesis and that of other components. For example, in the models developed by Scott {\it et al.}\cite{scott2010interdependence} and Maitra {\it et al.}\cite{maitra2015bacterial}, the active proteins correspond to ribosomes.  This class of models involving catalytic proteins can be used to accurately describe the exponential growth of a cell under a sufficient supply of substrates; however, once the degradation rate of the active protein exceeds its rate of synthesis under a limited substrate supply, the cell's volume will shrink, leading to cell death. Hence, a cell population either grows exponentially or dies out, and in this cellular state it is not possible to maintain the population without growth.\\
~~~~To model a state with such suppressed growth, we consider two more chemical species, inhibitors($I$) and active protein-inhibitor complexes($C$), in addition to the substrates($S$) and active proteins($P$) that are commonly adopted in models of cell growth. A schematic representation of the present model is shown in Fig.\ref{fig:Fig1}.(A). Here, we focus on two classes of proteins that are essential to the description of cellular growth: an active protein and inhibitor. The active proteins are those that catalyze their own growth such as ribosomes, and can include metabolic enzymes, transporters, and growth-facilitating factors. Inhibitory proteins form a complex with active proteins, thereby suppressing their catalytic synthesis function. They can be inhibitory factors such as YfiA or HPF in {\it Escherichia coli}. Other candidates for such inhibitors are misfolded or mistranslated proteins that are produced erroneously during the replication of active proteins, which inhibit the catalytic activity of active proteins by trapping them into the aggregates of misfolded proteins \cite{tyedmers2010cellular,oguchi2012tightly}. Our model, then, is given by
\begin{eqnarray}
\frac{dS}{dt}&=&-F_P(S)P-F_I(S)P+P(S_{\rm ext}-S)-\mu S \nonumber \\
\frac{dP}{dt}&=&F_P(S)P-G(P,I,C)-d_PP-\mu P \nonumber \\
\frac{dI}{dt}&=&F_I(S)P-G(P,I,C)-d_II-\mu I \label{eq:original}\\
\frac{dC}{dt}&=&G(P,I,C)-d_CC-\mu C \nonumber
\end{eqnarray}
 where $F_P(S)$ and $F_I(S)$ represent the synthesis rate of the active protein and inhibitor, to be presented below. $G(P,I,C)$ denotes the reaction of complex formation, given by $k_pPI-k_mC$, $S_{\rm ext}$ represents the external concentration of substrate, $d_i$ denotes the spontaneous degradation rate of macromolecules $i\  (i=P,I,C)$, and $\mu$ represents the specific growth rate of the cell, given by $\mu= F_P(S)P$.\\
~~~~In this model, the cell takes up substrates from the external environment from which active proteins and inhibitors are synthesized. These syntheses $\sext\leftrightarrow S,S\rightarrow P$, and $S\rightarrow I$, as well as the uptake of substrates, take place with the aid of catalysis by the active proteins. Then, by assuming that the synthesized components are used for growth in a sufficiently rapid period, the growth rate is set to be proportional to the rate of active protein synthesis. Next, the catalytic activity of the active protein is inactivated due to the formation of an active protein-inhibitor complex $P+I\leftrightarrow C$, which, for example, corresponds to the interaction between a ribosome and YfiA and HPF \cite{ueta2008role,vila2004structural,maki2000two}.  All chemical components are diluted by the volume growth of a cell, although they are spontaneously degraded at a much lower rate. The complex has higher stability than active protein and inhibitor, alone ($d_C$ is smaller than $d_P$ and $d_I$).\\
~~~~It has been well established that inhibitory factors are actively synthesized under a resource-limited condition \cite{jin2012growth}. Moreover, interpreting $I$ molecules as incorrect polymers, they are expected to increase with the decrease of supply substrates, since this situation will limit the proofreading mechanism to eliminate them\cite{hopfield1974kinetic}.  Thus, with this interpretation of the inhibitors as incorrect polymers and also in consistency with increase in inhibitory factors under resource-limited condition, it naturally follows that the ratio of the synthesis of active protein to inhibitors is an increasing function of substrate concentration, i.e., $\frac{d}{dS}\frac{F_P(S)}{F_I(S)}>0$.  In the model, we assume that this ratio increases with the concentration and becomes saturated at higher concentrations, as in Michaelis-Menenten's form, and choose $F_P(S)=\frac{vS}{K+S}\frac{S}{K_t+S}$ and $F_I(S)=\frac{vS}{K+S}\frac{K_t}{K_t+S}$, for example.  (see also the {\it Supplementary Information} for the derivation of such form in the case of a proofreading mechanism).  \\
~~~~Note that by summing up $\dot{P}$ and $\dot{C}$, we obtain $\dot{P}+\dot{C}=F_P(S)P(1-(P+C))$ if $d_P$ and $d_C$ are zero (or  negligible). It means that if the cell once reaches any steady state, the relationship $P+C=1$ is kept satisfied as long as $P$ and $F_P(S)$ are not zero. We use the relationship and eliminate $C$ by substituting $C=1-P$ for analysis below. 

\section*{Growth phases}
~~~~The steady state of the present model exhibits three distinct phases as a function of the external substrate concentration $\sext$ (Fig.\ref{fig:Fig1}.(B)), as computed by its steady-state solution.
The three phases are distinguished by both the steady growth rate and the concentration of active protein, which are termed as the active, inactive, and death phases, as shown in the figure, whereas the growth rate shows a steep jump at the boundaries of the phases. The phases are characterized as follows. {\bf (i)} In the active phase, the highest growth rate is achieved, where abundant active proteins work freely as catalysts. {\bf (ii)} In the inactive phase, growth rate is not zero but is drastically reduced with orders of magnitude compared with the active phase. Here, almost all active proteins are arrested by complex formation with the inhibitor, and their catalytic activity is deactivated. {\bf (iii)} At the death phase, a cell cannot grow, and all of the active proteins, inhibitors, and complexes go to zero. In this case, the cell goes beyond the so-called "point of no return" and can never grow again, regardless of the amount of increase in $\sext$, since the catalysts are absent in any form. (As will be shown below, the active and inactive phases correspond to the classic log and stationary phases, but to emphasize the single-cell growth mode, we adopt these former terms for now).\\
~~~~The transition from the active to inactive phase is caused by the interaction between the active protein and inhibitor. In the substrate-poor condition, the amount of inhibitor greatly exceeds the total amount of catalytic proteins ($P+C$), and any free active protein remaining vanishes. Below the transition point from the inactive to death phase, the spontaneous degradation rate surpasses the synthesis rate, at which point all of the components decrease. This transition point is simply determined by the balance condition $F_P=d_P$. Hence, if $d_P$ is set to zero, the inactive-death transition does not occur.\\
~~~~We now consider the time series of biomass (the total amount of macromolecules) that is almost proportional to the total cell number, under a condition with a given finite resource, for  comparison with experimental data in the batch culture condition (Fig.\ref{fig:Fig1}.(C and D)).  In the numerical simulation, the condition with a given, finite amount of substrates corresponding to the increase of cell number is implemented by introducing the dynamics of external substrate concentration to the original model. Here, $\sext$ is decreased as the substrates are replaced by the biomass, resulting in cell growth (details are given in the {\it Supplementary Information}). At the beginning of the simulation, the amount of biomass (i.e., cell number) stays almost constant, and then gradually starts increases exponentially. After the phase of exponential growth, substrates are consumed, and the biomass increase stops. Then, over a long time span, the biomass stays at a nearly constant value, until it begins to slowly decrease. Finally, the degradation dominates and the biomass (cell number) falls off dramatically. \\
~~~~These successive transitions in the growth of biomass (Fig.\ref{fig:Fig1}(C and D)) from initially inactive to the active, inactive, and death phases corresponds to those among the lag, log, stationary, and death phases. As the initial condition was chosen as the inactive phase under a condition of rich substrate availability, most of the active proteins are arrested in a complex at this point. Therefore, at the initial stage, dissociation of the complex into active proteins and inhibitors progress, and biomass is barely synthesized, even though rich substrate is available. After the cell escapes this waiting mode, catalytic reactions from active proteins progress, leading to an exponential increase in biomass. Subsequently, the external substrate is depleted, and cells experience another transition from the active to inactive phase. At this point, the biomass decreases only slowly owing to the remaining substrate and stability of the active protein-inhibitor complex. However, after the substrate is depleted and the active protein and inhibitor are dissociated from the complex, the biomass decreases at a much faster rate, ultimately entering the death phase.\\
~~~~In the active phase with exponential growth, the present model exhibits classical growth laws, namely {\bf (i)} Monod's growth law, {\bf (ii)} Pirt's law, and {\bf (iii)} growth rate vs. ribosome fraction (see {\it Supplementary Information} Fig. S1).\\

\section*{Lag time dependency on starvation time $\bvec{\tst}$ and maximum growth rate $\bvec{\mu_{\rm max}}$}
~~~~In this section, we uncover the quantitative relationships among the basic quantities characterizing the transition between the active and inactive phases; i.e., lag time, starvation time, and growth rates. We demonstrate that the theoretical predictions agree well with experimentally observed relationships.
\\
~~~~First, we compute the dependency of lag time $(\lambda)$ on starvation time $(\tst)$ and the maximum specific growth rate $(\mu_{\rm max})$.
Up to time $t=0$, the model cell is set in a substrate-rich condition $\sext=\srich$, and stays at a steady state with exponential growth. Then, the external substrate is depleted to $\sext=\spoor$ instantaneously. The cell is exposed to this starvation condition up to starvation time $t=\tst$.  Subsequently, the substrate concentration $\sext$ instantaneously returns to $\srich$. After the substrate level is recovered, it takes a certain length of time for a cell to return to its original growth rate (Fig. S2), which is the lag time $\lambda$ (following the standard definition of introduced by Penfold and Pirt\cite{penfold1914nature,pirt1975principles}). Given this, the dependency of $\lambda$ on the starvation time $\tst$ and $\mu_{\rm max}$ can be computed.  \\
\subsection*{Relationship between lag and starvation time: $\bvec{\lambda\propto\sqrt{\tst}}$}
\begin{figure}[h!]
\centering
\includegraphics[width = 120mm, angle = 0,bb=0 0 1153 981]{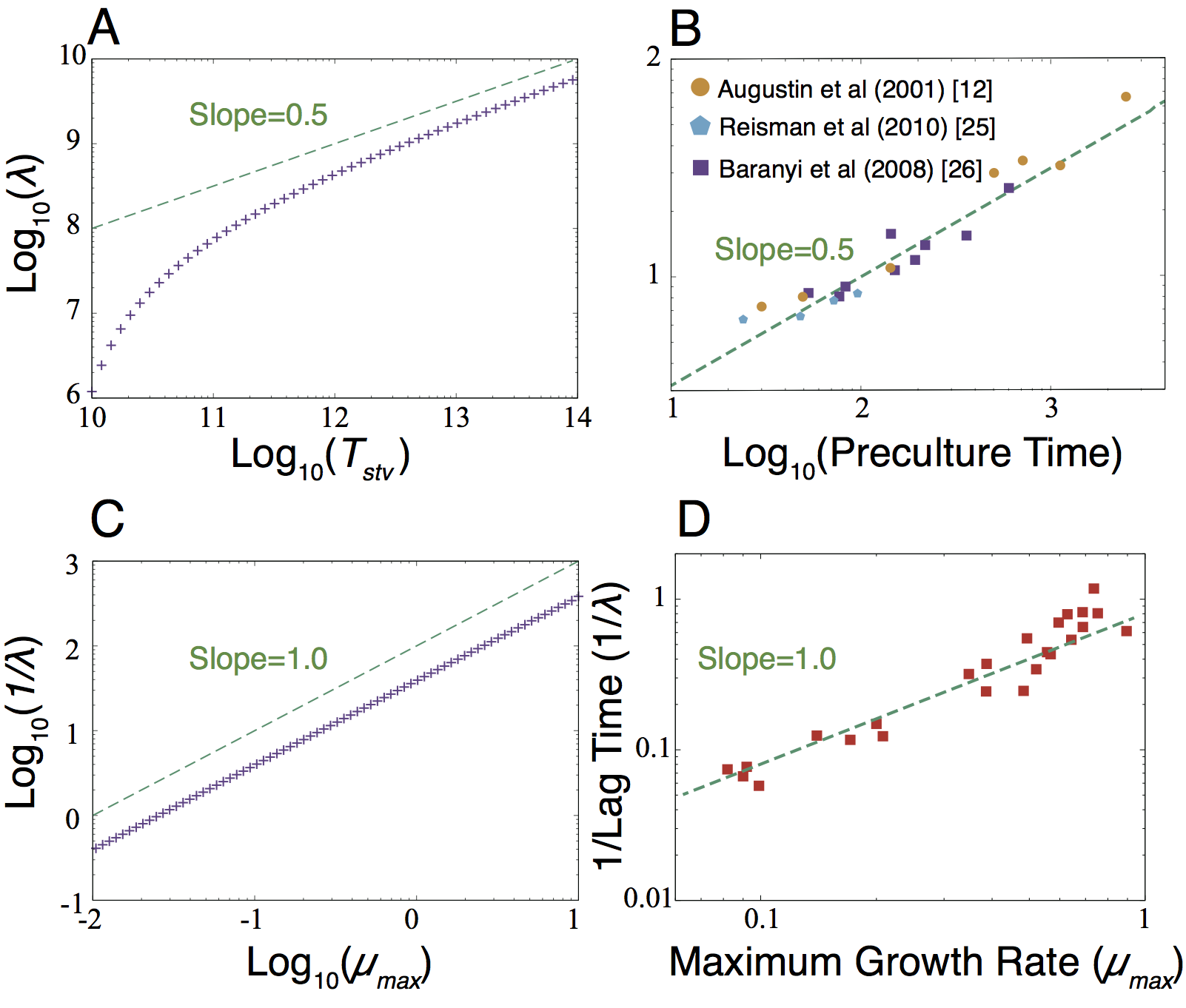}
     \caption{(A and B) Lag time is plotted as a function of (A) starvation time or (B) pre-incubation time. (C and D) Relationship between the lag time and maximum specific growth rate $\mu_{\rm max}$.  Parameters were set as follows: $\srich = 10^8,\spoor=10^{-1}$, same parameter values as Fig.\ref{fig:Fig1} were adopted except $v=0.1$ and $d_i=0$ here. Data are adopted from \cite{levin2010automated,pin2008single,augustin2000model} and \cite{oscar2005validation} for (B) and (D), respectively. Absolute values of lag time are rescaled among data. Lag time is computed as the time needed to reach the steady state under $\sext=\srich$ from an initial condition in inactive phase. In (C), it is obtained by varying $v (=\mu_{\rm max})$.}
    \label{fig:Fig2}
\end{figure}

~~~~We found that $\lambda$ increases proportionately to $\sqrt{\tst}$, as shown in Fig.\ref{fig:Fig2}(A). The experimentally observed relationship between $\lambda$ and $\tst$ is also plotted for comparison in Fig.\ref{fig:Fig2}(B), using reported data \cite{levin2010automated,pin2008single,augustin2000model}, which also exhibited $\lambda\propto\sqrt{\tst}$ dependency. Although this empirical dependency has been previously discussed, its theoretical origin has thus far not been uncovered. \\
~~~~Indeed, the origin of $\lambda \propto\sqrt{\tst}$ is explained by noting the anomalous relaxation of inhibitor concentration, which is caused by the interaction between the active protein and inhibitor.  The sketch of this explanation is given below, and the analytic derivation is given in the {\it Supplementary Information}.\\
~~~~First, consider the time course of chemical concentrations during starvation. In this condition, cell growth is inhibited by two factors: substrate depletion and deactivation of catalytic activity of the active protein. Following the decrease in uptake due to depletion of $\sext$, the concentration of $S$ decreases, resulting in a change in the balance between $P$ and $I$. (Hereafter we adopt the notation such that $P$, $I$, and $C$ also denote the concentrations of corresponding chemicals). Under the $\spoor$ condition, the ratio of the synthesis of $I$ to $P$ increases. With an increase in $I$, $P$ decreases due to the formation of a complex with $I$. Over time, more $P$ gets arrested, and the level of inactivation increases with the duration of starvation. \\
~~~~In this scenario, the increase of concentration $I$ is slow. Considering that the complex formation reaction $P+I\leftrightarrow C$ rapidly approaches its equilibrium, i.e., $k_pPI\sim k_mC$,  $P$ is roughly proportional to the inverse of $I$ (recall $P+C=1$), if $I$ is sufficiently large. Accordingly, the synthesis rate of $I$, given by $F_I(S)P$, is inversely proportional to its amount, i.e., $\dot{I}(t)\propto F_I(S)/I,$ and thus $\bvec{dI^2/dt}{\bf\sim const..}$ Hence, the inhibitor accumulation progresses with  $I(t)\propto \sqrt{t}$. (Note that due to $S$ depletion, the dilution effect is negligible.)\\
~~~~Next, we consider the time course for the resurrection after recovery of the external substrate. During resurrection, $P$ is increased while $I$ is reduced. Since $P$ is strongly deactivated after starvation, the dilution effect from cell growth is the only factor contributing to the reduction of $I$. Noting $\mu=F_PP$ and $P\propto1/I$, the dilution effect is given by $\mu I=F_PPI\propto I/I=const.$ at the early stage of resurrection. Thus, the resurrection time course of $I$ is determined by the dynamics
$\bvec{\dot{I}(t)}{\bf \propto -const.,}$
leading to the linear decrease of $I$, i.e., $I(t)\sim I(0) -const. \times t$.\\
~~~~Let us briefly recapitulate the argument presented so far. The accumulated amount of component $I$ is proportional to $\sqrt{\tst}$, while during resurrection, the dilution of $I$ progresses linearly with time, which is required for the dissociation of $P$ and $I$, leading to growth recovery. By combining these two estimates, the lag time satisfies $\lambda\propto\sqrt{\tst}$.
\subsection*{Relationship between the lag time and maximal growth rate: $\bvec{\lambda \propto1/\mu_{\rm max}}$}
 ~~~~Second, the relationship $\lambda\propto 1/\mu_{\rm max}$ is obtained by numerical simulation of our model, in line with experimental results \cite{oscar2005validation} (Fig.\ref{fig:Fig2}(C and D)).\\
~~~~This relationship $\lambda\propto1/\mu_{\rm max}$ is also explained by the characteristics of the resurrection time course. The dilution rate of $I$ over time is given by  $\mu I$, as mentioned above; thus, at the early stage, $\dot{I}\sim -\mu I$. In the substrate-rich condition, the substrate abundances are assumed to be saturated, so that $$\lim_{\srich\to \infty}\dot{I}\sim\lim_{\srich\to\infty}F_P\cdot I/I= \mu_{\rm max}$$ holds because $\lim_{S\to\infty} F_P(S)=\mu_{\rm max}$ is satisfied. Thus, it follows that $\lambda\propto 1/\mu_{\rm max}$.\\
~~~~We also obtained an analytic estimation of the lag time as
\begin{equation}
\lambda\sim \frac{1}{\mu_{\rm max}}\sqrt{2F_IK_A\tst}, \label{eq:laggrowth}
\end{equation}
where $K_A=k_p/k_m$ (see {\it Supplementary Information} for conditions and calculation). In this form, the two relationships $\lambda\propto \sqrt{\tst}$ and $\lambda \propto 1/\mu_{\rm max}$ are integrated.\\
The present theory also explains other experimental observations. First, in predictive microbiology\cite{swinnen2004predictive}, the lag time to return the log phase from the stationary phase is regarded as the time span required to consume the $<work>$ accumulated during the stationary phase with the rate $\mu_{\rm max}$. Thus, the amount of $<work>$ is defined as the product of $\lambda$ and $\mu_{\rm max}$.
In our results, accumulated inhibitor $I$ needs to be consumed during the lag time $\lambda$, so that $<work>$ is interpreted as $I$, whose time course agrees well with that of $<work>$ obtained experimentally (See Supplementary figure Fig.S4). Second, the tradeoff between the growth rate and tolerance for the starvation, experimentally observed \cite{vasi1999ecological} is also derived from our theory (See {\it Supplementary Information}).
\section*{Dependence of lag time on the starvation process}
\begin{figure}[h!]
\centering
\includegraphics[width = 120 mm, angle = 0,bb=0 0 895 418]{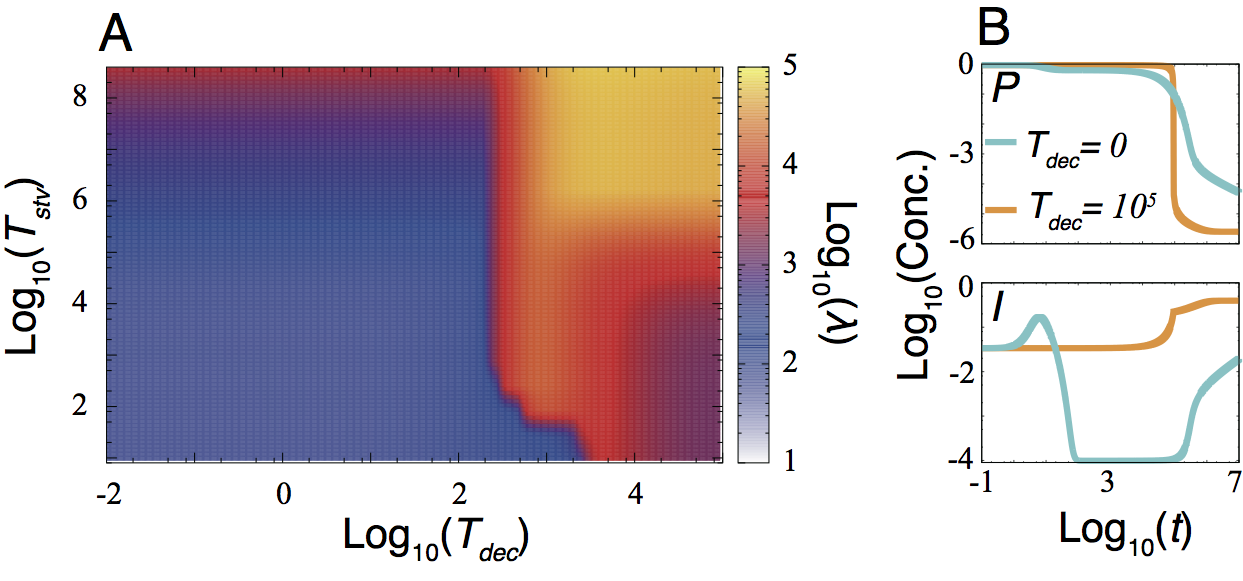}
     \caption{(A) Dependence of lag time $\lambda$ on the time required to decrease the substrate $T_{\rm dec}$ and starvation time $T_{\rm stv}$. (B) Time series of relaxation (responses to decrease of the substrate) of the active protein $P$, and inhibitor $I$. The orange line is the orbit of a slow substrate decrease ($T_{\rm dec}=10^5$), and the cyan line indicates the time course of an instantaneous substrate decrease ($T_{\rm dec}=0$).  Parameters were set as follows: $ v=0.1,K=5.0,K_t=100.0,k_p=0.2,k_m=2\times 10^{-7},d_R=d_I=d_C=0,\srich=10^{3},\spoor=10^{-3}$ and $T_{\rm stv}=10^7$.}
    \label{fig:Fig3}
\end{figure}

\begin{figure}[h!]
\centering
\includegraphics[width = 120 mm, angle = 0,bb=0 0 1045 898]{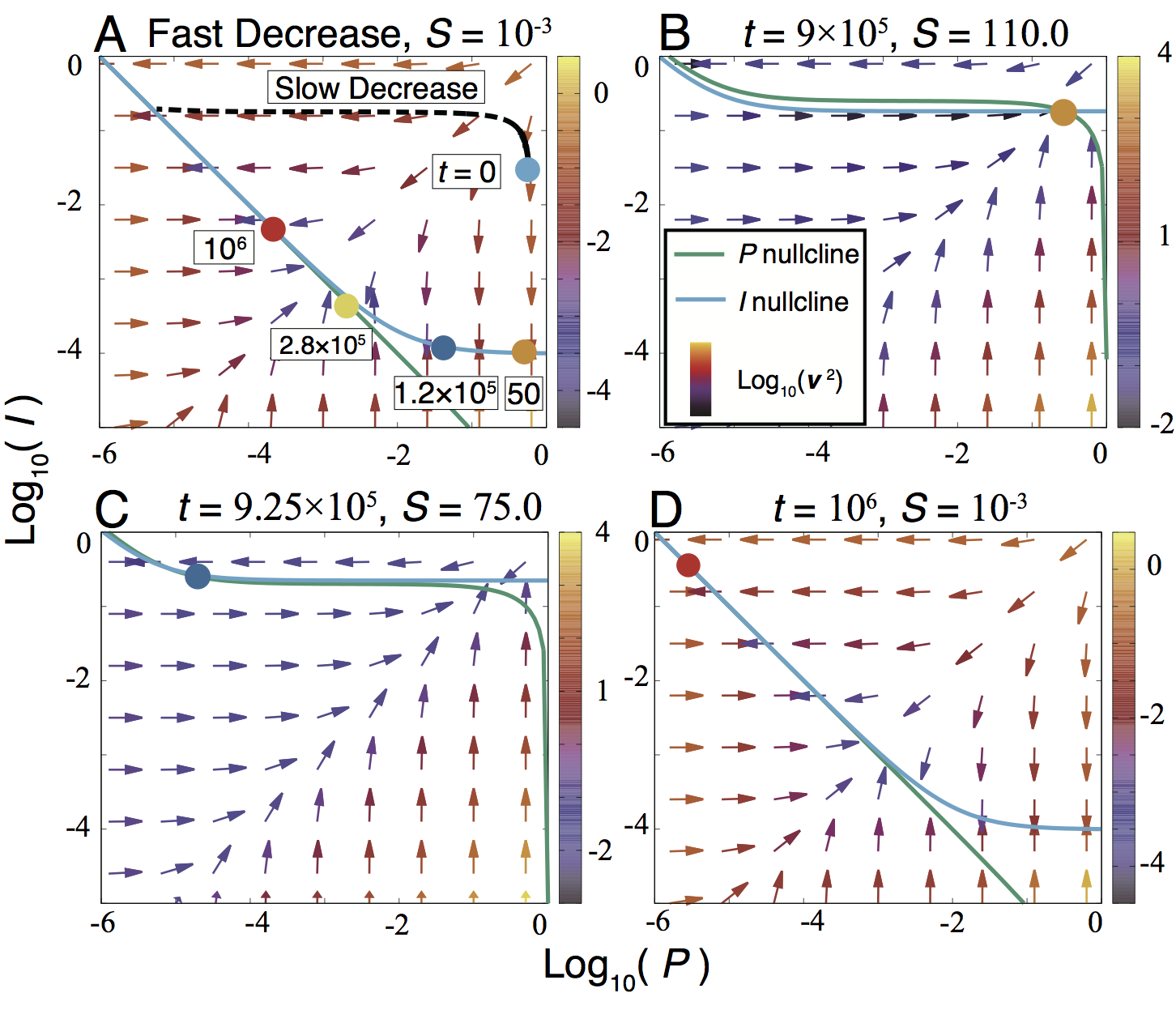}
     \caption{Movement of nullclines (skyblue lines) and time evolution of state variables (circles with in the state space $(P,I)$. (A) The case of fast substrate decrease (number in white boxes indicate the time point). The orbit of slow substrate decrease is also plotted. (B-D) The case of slow substrate decrease.  Each point is the value of the state variable at the indicated time and substrate concentration. The vector field $\bvec{v}=(dP/dt,dI/dt)$ is also depicted. Parameters are identical with that described in Fig.\ref{fig:Fig3} }
    \label{fig:Fig4}
\end{figure}

~~~~So far, we have considered the dependence of lag time on the starvation time. However, in addition to the starvation period, the starvation process itself, i.e., the speed required to reduce the external substrate, has an influence on the lag time.\\
~~~~For this investigation, instead of the instantaneous depletion of the external substrate, its concentration is instead gradually decreased over time in a linear manner over the span $\tdec$, in contrast to the previous simulation procedure, which corresponds to $\tdec=0$.
Then, the cell is placed under the substrate-poor condition for the duration $\tst$, before the substrate is recovered, and the lag time $\lambda$ is computed.\\
~~~~The dependence of the lag time $\lambda$ on $\tst$ and $\tdec$ is shown in Fig.\ref{fig:Fig3}(A). While $\lambda$ monotonically increases against $\tst$ for a given $\tdec$, it shows drastic dependence on $\tdec$. If the external concentration of the substrate is reduced quickly (i.e., small $\tdec$), the lag time is rather small. However, if the decrease in the external substrate concentration is slow (i.e., large $\tdec$), the lag time is much longer. In addition, this transition from a short to long lag time is quite steep. \\
~~~~This transition against the timescale of the environmental change manifests itself in the time course of chemical concentrations (see Fig.\ref{fig:Fig3}(B)). With rapid environmental change, $S$ decreases first, whereas with slow environmental change, $P$ decreases first. In addition, the value of $I$ is quite different between the two cases, indicating that the speed of environmental change affects the degree of inhibition, i.e., the extent to which active proteins are arrested by inhibitors to form a complex. \\
~~~~Now, we provide an intuitive explanation for two distinct inhibition processes. When $\sext$ starts to decrease, a cell is in the active phase in which $P$ is abundant. If the environment changes sufficiently quickly, there is not enough time to synthesize the chemicals $P$ or $I$, because of the lack of $S$, and the concentrations of chemical species are frozen near the initial state with abundant $P$. However, if the environmental change is slower than the rate of the chemical reaction, the concentration of the inhibitor $I$ (active protein $P$) increases (decreases), respectively. Hence, $P$ remains rich in the case of fast environmental change, whereas $I$ is rich for a slow environmental change.  In the former case, when the substrate is increased again, the active proteins are ready to work, so that the lag time is short, which can be interpreted as a kind of "freeze dry" process. Note that the difference in chemical concentration caused by different $\tdec$ is maintained for log time because in slow (fast) environmental change, chemical reactions are almost halted due to the decrease of $P$ ($S$), respectively. Thus, the difference of lag time remains even for large $\tst$ as Fig.\ref{fig:Fig3}(A) (The mechanism of this slow process is discussed in {\it Supplementary Information}.).\\
~~~~This lag time difference can also be explained from the perspective of dynamical systems\cite{strogatz2014nonlinear}. For a given $S$, the temporal evolution of $P$ and $I$ is given by the flow in the state space of $(P,I)$.  Examples of the flow are given in Fig.\ref{fig:Fig4}.  The flow depicts $(dP/dt,dI/dt)$, which determines the temporal evolution. The flow is characterized by $P-$ and $I-$ nullclines, which are given by the curves satisfying $dP/dt=0$ and $dI/dt=0$, as plotted in Fig.\ref{fig:Fig4}. \\
~~~~Note that at a nullcline, the temporal change of one state variable (either $P$ or $I$) vanishes. Thus, if two nullclines approach each other, then the time evolution of both the concentrations $P$ and $I$ are slowed down, and the point where two nullclines intersect corresponds to the steady state. As shown in Fig. \ref{fig:Fig4}, nullclines come close together under the substrate-depleting condition, which gives a dynamical systems account of the slow process in the inactive phase discussed so far. \\
~~~~For a fast change (i.e., small $\tdec$, Fig.\ref{fig:Fig4}(A)), $S$ is quickly reduced at the point where the two nullclines come close together. Then, the dynamics of $(P,I)$ follow the flow as shown in the figure. First, $I$ decreases to reach the $I$-nullcline. Then, the state changes along the almost coalesced nullclines when the dynamics are slowed down. Thus, it takes a long time to decrease the $P$ concentration, so that at resumption of the substrate, sufficient $P$ can be utilized. \\
~~~~In contrast, for a slow change (i.e., large $\tdec$), the flow in $(P,I)$ gradually changes as shown in Fig.\ref{fig:Fig4}(B-D). Initially, the state $(P,I)$ stays at the substrate-rich steady state (Fig.\ref{fig:Fig4}(B)). Due to the change in substrate concentration, two nullclines moderately move and interchange their vertical locations. Since the movement of nullclines is slow, the decrease in $P$ progresses before the two nullclines come close together (i.e., before the process is slowed down). The temporal evolution of $P$ and $I$ is slowed down only after this decrease in $P$(Fig.\ref{fig:Fig4}(C and D)). Hence, the difference between the cases with small and large $\tdec$ is determined by whether the nullclines almost coalesce before or after the $P$ decrease, respectively. 
\section*{Distribution of lag time}
\begin{figure}[h!]
\centering
\includegraphics[width = 120 mm, angle = 0,bb=0 0 970 708]{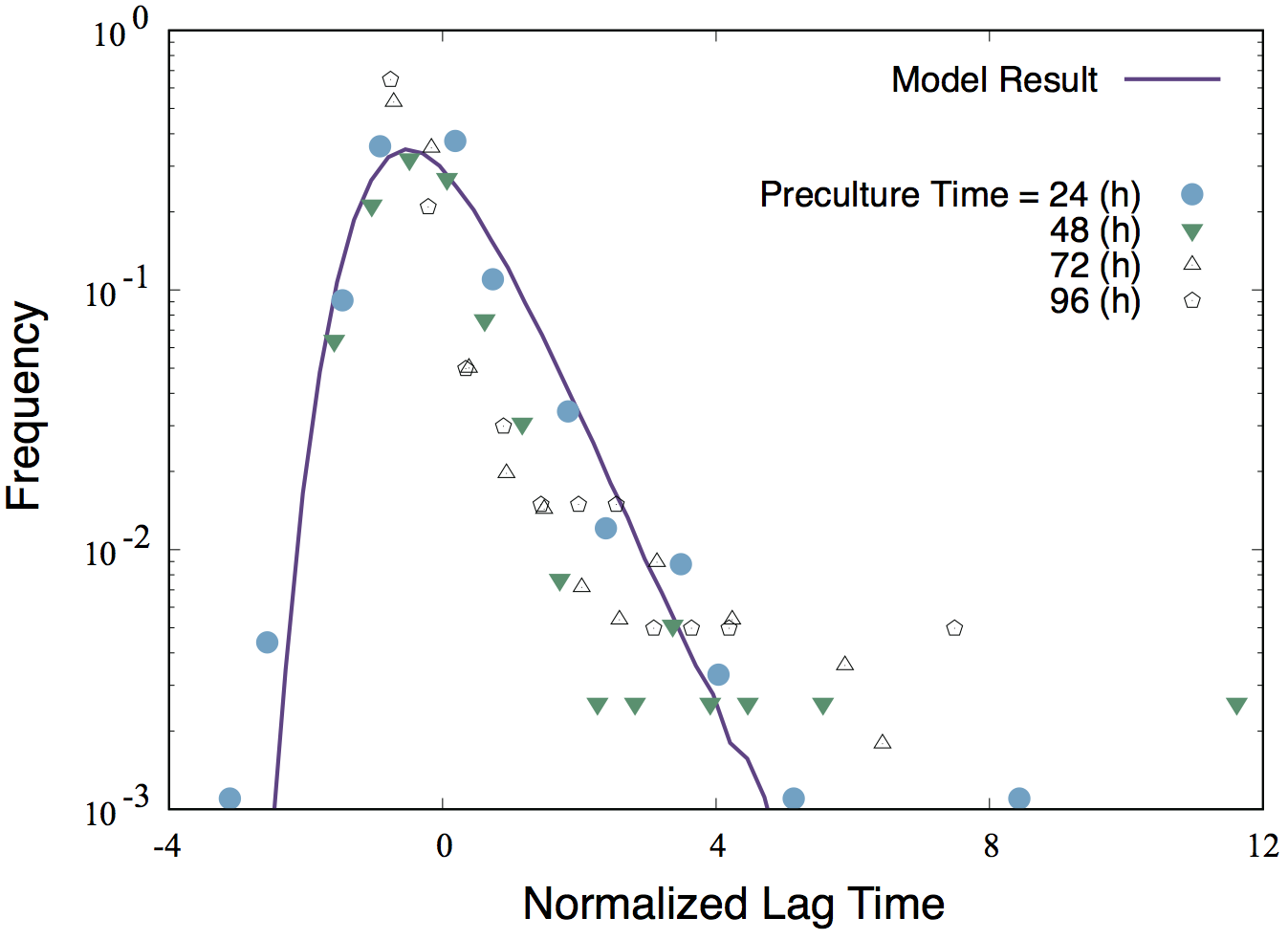}
     \caption{Distribution of lag time obtained by model simulation (solid line) with experimental results overlaid. The horizontal axis of each distribution is normalized using its mean $E$ and standard deviation $\sigma$ as $\lambda\to (\lambda-E)/\sigma$. The experimental data are adapted from \cite{levin2010automated}. Methods of stochastic simulations and parameter values are given in {\it Supplementary\ Information}.}
    \label{fig:Fig5}
\end{figure}

~~~~So far, we have considered the average change of chemical concentrations using the rate equation of chemical reactions. However, the biochemical reaction is inherently stochastic, and thus the lag time is accordingly distributed. This distribution was computed by carrying out a stochastic simulation of chemical kinetics using the Gillespie algorithm\cite{gillespie1977exact}. \\
~~~~We found that the distribution of lag time $\lambda$ has a standard Gaussian form for the shorter lag-time side but has an $\exp(-\lambda)$ tail for the longer side (Fig.\ref{fig:Fig5}). This exponential tail was also observed in experiments, as overlaid in Fig.\ref{fig:Fig5}, which is adapted from Reisman {\it et al.} \cite{levin2010automated}. In the present model, once the number of active proteins becomes small, more time is needed to recover the growth, so that the distribution of initial active protein abundances is expanded to a long-tailed distribution. The agreement of the model with experimental data is relatively good for a short starvation time (24 hours and 48 hours) but for longer times, the experimental data may suggest the existence of a much longer tail.\\

\section*{Summary and Discussion}
~~~~We developed a coarse-grained model consisting of a substrate, autocatalytic active protein, inhibitor of the active protein, and active protein-inhibitor complex. In the steady state, the model shows distinct phases, i.e., active, inactive, and death phases. In addition, the temporal evolution of total biomass shows bacterial growth curve-like behavior. The present model is not only consistent with the already-known growth laws in the active phase but also demonstrates two relationships, $\lambda\propto \sqrt{\tst}$ and $\lambda\propto 1/\mu_{\rm max}$, concerning the duration of the lag time $\lambda$. Although these two relationships have also been observed experimentally, their origins and underlying mechanisms had not yet been elucidated. The present model can explain these relationships based on the formation of a complex between the active protein and inhibitor, whose increase in the starvation condition hinders the catalytic reaction. The inactive phase, which corresponds to the stationary phase, as well as the above two laws are generally derived as long as the ratio of the synthesis of the inhibitor to that of the active protein is increased along with a decrease in the external substrate concentration.  This condition is also derived if the inhibitor is interpreted as a product of erroneous protein synthesis, where a proofreading mechanism to correct the error needing energy works inefficiently in a substrate-poor condition. \\
~~~~Although the cell state with exponential growth has been extensively analyzed in previous theoretical models, the transition to the phase with suppressed growth has thus far not been theoretically explained. Our model, albeit simple, provides an essential mechanism for this transition as complex formation of active and inhibitor proteins, which can be experimentally tested.\\
~~~~Moreover, the model predicts that the lag time differs depending on the rate of external depletion of the substrate, which can also be examined experimentally. Recently, the bimodal distribution of growth resumption time from the stationary phase was reported in a batch culture experiment\cite{joers2016growth}. The heterogeneous depletion of a substrate due to the spatial structure of a bacterial colony is thought to be a potent cause of this bimodality, while understanding of this concept is fairly underway. Since the present model shows different lag times for different rates of environmental change, it can provide a possible scenario for explaining this bimodality.\\

\section*{Acknowledge}The authors would like to thank S.Krishna, S.Semsey, N.Mitarai,  A.Kamimura, N.Saito, and T. S. Hatakeyama for useful discussions; and I. L. Reisman, N. Balaban, and J.C. Augustin for providing data.  This research is partially supported by the Platform for Dynamic Approaches to Living System from Japan Agency for Medical Research and Development (AMED), Grant-in-Aid for Scientific Research (S) (15H05746 from JSPS), and the Japan Society for the Promotion of Science(16J10031).

\end{document}